\documentclass[
  aps,
  pra,
  superscriptaddress,
  a4paper,
notitlepage,
reprint,
longbibliography
]{revtex4-1}
\usepackage{amsmath}
\usepackage{amssymb}
\usepackage{bm}
\usepackage{graphicx}
\usepackage{microtype}

\usepackage[hidelinks]{hyperref}

\newcommand*\diff{\mathop{}\!\mathrm{d}}
\newcommand*\ee{e}
\newcommand*\ii{i}
\newcommand*\abs[1]{\lvert#1\rvert}

\DeclareMathOperator\real{Re}
\DeclareMathOperator\imag{Im}
\DeclareMathOperator\tr{Tr}
\def\<#1>{\mathinner{\langle#1\rangle}}

\ifdefined\pdfvariable
  \def\pdfpageattr{\pdfvariable pageattr}
\fi
\pdfpageattr{/Group <</S /Transparency /I true /CS /DeviceRGB>>}

\begin{document}

\title{Bogoliubov Fermi surfaces stabilized by spin-orbit coupling}

\author{Henri Menke}
\email{henrimenke@gmail.com}
\affiliation{Department of Physics and MacDiarmid Institute for
  Advanced Materials and Nanotechnology, University of Otago, P.O. Box
  56, Dunedin 9054, New Zealand}
\author{C. Timm}
\email{carsten.timm@tu-dresden.de}
\affiliation{Institute of Theoretical Physics, Technische
  Universit\"at Dresden, 01062 Dresden, Germany}
\author{P. M. R. Brydon}
\email{philip.brydon@otago.ac.nz}
\affiliation{Department of Physics and MacDiarmid Institute for
  Advanced Materials and Nanotechnology, University of Otago, P.O. Box
  56, Dunedin 9054, New Zealand}

\date{\today}

\begin{abstract}
  It was recently understood that centrosymmetric multiband
  superconductors that break time-reversal symmetry generically show
  Fermi surfaces of Bogoliubov quasiparticles.  We investigate the
  thermodynamic stability of these Bogoliubov Fermi surfaces in a
  paradigmatic model. To that end, we construct the mean-field phase
  diagram as a function of spin-orbit coupling and temperature. It
  confirms the prediction that a pairing state with Bogoliubov Fermi
  surfaces can be stabilized at moderate spin-orbit coupling
  strengths. The multiband nature of the model also gives rise to a
  first-order phase transition, which can be explained by the
  competition of intra- and interband pairing and is strongly affected
  by cubic anisotropy.  For the state with Bogoliubov Fermi surfaces,
  we also discuss experimental signatures in terms of the residual
  density of states and the induced magnetic order.  Our results show
  that Bogoliubov Fermi surfaces of experimentally relevant size can
  be thermodynamically stable.
\end{abstract}

\maketitle

\section{Introduction}

A hallmark of unconventional superconductivity is a nodal pairing
state where the excitation gap vanishes at points or lines in
momentum space~\cite{Mineev1999}.  Recently, however, a third type of
node has been proposed: extended Bogoliubov Fermi surfaces (BFSs)
where the excitation gap vanishes at a surface in momentum space
\cite{Agterberg2017, Brydon2018b}.  In clean, inversion-symmetric
(even-parity) superconductors that spontaneously break time-reversal
symmetry (TRS), all nodes are generically expected to be BFSs.
Crucial for the appearance of BFSs is that the superconductivity
involves more than one band: Specifically, the pairing between
electrons in different bands generates a pseudomagnetic field, which
``inflates'' point and line nodes of the intraband pairing potential
into BFSs. These nodal surfaces are robust against perturbations that
preserve particle-hole and inversion symmetries, which can be formulated
in terms of a $\mathbb{Z}_2$ topological
invariant~\cite{Kobayashi2014, Zhao2016, Agterberg2017, Bzdusek2017}.

A natural setting for the appearance of BFSs is in systems where a
multiband structure arises from the presence of discrete low-energy
electronic degrees of freedom apart from spin, e.g., atomic-orbital or
sublattice indices. This permits the construction of novel
``internally anisotropic'' pairing states where the Cooper-pair
wave function has nontrivial dependence upon the orbital or sublattice
indices~\cite{Nomoto2016b, Cheung2018, Brydon2018b}. Crucially, for the
appearance of BFSs, internally anisotropic pairing states are
typically characterized by both intraband and interband pairing
potentials~\cite{Brydon2018b}. Such pairing states have been proposed
for a wide variety of multiband systems of current interest, such as
iron-based superconductors~\cite{Gao2010, Nicholson2012, Ong2016,
  Vafek2016, AgterbergFeSe2017, Setty2019, Hu2019},
Cu$_x$Bi$_2$Se$_3$~\cite{Fu2014}, half-Heusler
compounds~\cite{Brydon2016, Agterberg2017, Savary2017, Yang2017,
  Roy2017, Timm2017, Boettcher2018a, Boettcher2018b, Kim2018,
  Brydon2018b}, the antiperovskite Sr$_{3-x}$SnO~\cite{Kawakami2018},
Sr$_2$RuO$_4$~\cite{Taylor2012}, UPt$_3$~\cite{Nomoto2016a,
  Yanase2016}, transition-metal dichalcogenides~\cite{Oiwa2018,
  Moeckli2018}, and twisted bilayer graphene~\cite{Guo2018, Su2018,
  FWu2019}. This long list of materials---some of which are believed
to support a time-reversal-symmetry-breaking (TRSB) state---is
encouraging for the existence of~BFSs.

Although BFSs are robust against symmetry-preserving perturbations,
this topological protection does not guarantee the existence of such a
state.  Instead, it is necessary to consider the thermodynamic
stability. Since a TRSB combination of two nodal pairing states
eliminates all nodes that are not common to both states, it is expected
to be energetically favored over time-reversal-symmetric
combinations~\cite{Sigrist1991}. This argument does not hold if the
resulting TRSB state possesses a BFS, however, as this implies a
nonzero density of states (DOS) at the Fermi energy, which, at first
glance, is unfavorable compared to the line nodes generic for
time-reversal-symmetric states.  It was argued in Ref.~\cite{Agterberg2017}
that a TRSB state with a BFS could, nevertheless, be energetically
favorable in the presence of sufficiently strong spin-orbit coupling
(SOC). This analysis was restricted to temperatures close to $T_c$,
however, and so did not account for the effect of the expected large
residual DOS at low temperatures. Moreover, although the TRSB state
becomes more stable with increasing SOC, the size of the BFS
decreases, as shown below.  It is, thus, unclear if BFSs can be realized
in a limit where they have a detectable effect on the electronic
structure~\cite{Lapp2019}.  Another interesting question raised by the
analysis in Refs.~\cite{Agterberg2017, Brydon2018b} is what happens at SOC
strengths insufficient for a stable TRSB state.

In this paper, we use mean-field theory to study the appearance of
BFSs in a paradigmatic model of a multiband system with strong SOC,
specifically the Luttinger-Kohn Hamiltonian of $j=3/2$ fermions in a
cubic material~\cite{LK56}. The $j=3/2$ degree of freedom naturally
leads to a multiband system and to internally anisotropic
pairing. Assuming pairing in a \textit{s}-wave $J=2$
channel~\cite{Brydon2016}, we construct the superconducting phase
diagram as a function of the SOC strength and temperature.  We focus
on a particular set of pairing states belonging to the irreducible
representation (irrep) $T_{2g}$, which is expected to provide a
typical picture for pairing in a higher-dimensional representation. At
vanishing SOC, a fully gapped time-reversal-symmetric superconducting
phase is realized as was predicted in Ref.~\cite{Ho1999}.  For nonzero
SOC, we obtain a rich phase diagram, which, in particular, contains a
sizable region with TRSB superconductivity. The largest BFSs that we
find lead to a residual zero-temperature DOS at the Fermi energy of
approximately 20\% of the normal-state DOS, which should leave clear
signatures in thermodynamic measurements. We also verify the existence
of a subdominant magnetic order parameter which is induced by the TRSB
superconductivity.

Our paper is organized as follows: In Sec.~\ref{sec:mft}, we
introduce our microscopic model and outline the mean-field theory,
including a discussion of previously known limits of vanishing and
strong SOC. We present the mean-field phase diagram in
Sec.~\ref{sec:trsb} and study the effect of cubic anisotropy of the
SOC. A key feature of the phase diagram is the first-order transition
into a time-reversal-symmetric superconducting state at intermediate
SOC strength, which we explain in terms of a simplified model.  This
is followed in Sec.~\ref{sec:trsbstate} by a detailed study of the
TRSB state and the induced magnetic order parameter. We summarize our
results and draw additional conclusions in Sec.~\ref{sec:conclusion}.

\section{Model and mean-field theory
\label{sec:mft}}

Our starting point is the Luttinger-Kohn Hamiltonian for $j=3/2$
fermions in a cubic material~\cite{LK56},
\begin{equation}
  \label{eq:LK}
  h(\bm{k}) = (\alpha \abs{\bm{k}}^2 - \mu)\, \openone_4 + \beta \sum_i k_i^2 J_i^2 + \gamma \sum_{i \neq j} k_i k_j J_i J_j ,
\end{equation}
where $i = x,y,z$ and $i+1 = y$ if $i = x$, etc., and $J_i$ are the
$4\times 4$ matrix representations of the angular momentum operators
$j = 3/2$. The $j=3/2$ fermions can arise due to the strong atomic
SOC, e.g., of spins $s=1/2$ and orbital angular momenta $l=1$ for
\textit{p} orbitals.  In addition to the spin-independent dispersion
coefficient $\alpha$ and the chemical potential $\mu$, the Hamiltonian
in Eq.~\eqref{eq:LK} includes the symmetry-allowed SOC terms
proportional to $\beta$ and $\gamma$. The Hamiltonian has doubly
degenerate eigenvalues given by
\begin{align}
  \label{eq:EV}
  \epsilon_{\bm{k},\pm}
  & = \biggl( \alpha + \frac{5}{4}\, \beta \biggr) \abs{\bm{k}}^2 - \mu \nonumber \\
  & \quad{} \pm \beta \sqrt{\sum_i \biggl[ k_i^4 + \biggl( \frac{3\gamma^2}{\beta^2} - 1 \biggr)
    k_i^2 k_{i+1}^2 \biggr]} .
\end{align}
Note that SOC lifts the fourfold degeneracy of the $j=3/2$ manifold
away from the $\Gamma$ point. Due to the presence of time-reversal and
inversion symmetries, the bands remain doubly degenerate so that the
states in each band can be labeled by a pseudospin-$1/2$
index~\cite{Brydon2018b}.

The description in terms of an effective spin $j=3/2$ permits Cooper
pairs with total angular momentum $J = 0$ (singlet) and $J = 1$
(triplet), but also $J = 2$ (quintet) and $J = 3$
(septet)~\cite{Boettcher2016, Brydon2016, Savary2017, Roy2017,
  Yang2017, Venderbos2018, Boettcher2018a, Boettcher2018b, Kim2018,
  Sim2019}. Similar to singlets and triplets, the quintet and septet
pairings correspond to even- and odd-parity orbital wave functions,
respectively. In particular, this allows for a broader variety of
\textit{s}-wave pairing states: Besides the usual singlet, there are
five additional quintet states with on-site pairing.

Restricting ourselves to such local pairing states, the pairing
interaction has the general form
\begin{equation}
  H_{\text{pair}} = \sum_{j} \sum_l \sum_{l_i\in l}
    V_l\, b^\dagger_{l_i,j}b_{l_i,j} ,
\end{equation}
where $b^\dagger_{l_i,j}$ creates a Cooper pair at site $j$ in channel
$l_i$ belonging to the irrep $l$~\cite{Brydon2016}. There are three
irreps in the cubic $O_h$ point group which support \textit{s}-wave
pairing: The singlet state belongs to the one-dimensional $A_{1g}$
irrep, whereas the five quintet states are distributed into the
two-dimensional $E_g$ and the three-dimensional $T_{2g}$ irreps.

Within the standard mean-field treatment, the interaction is decoupled
to obtain the effective single-particle Hamiltonian,
\begin{equation}
  \label{eq:H_MF}
  H_{\mathrm{MF}} = \sum_{\bm{k}} \biggl(
  \frac{1}{2}\, \Psi_{\bm{k}}^\dagger \mathcal{H}(\bm{k}) \Psi_{\bm{k}}
  + \sum_l \frac{\tr[\Delta_l \Delta_l^\dagger]}{V_l} \biggr),
\end{equation}
with the Bogoliubov--de~Gennes (BdG) Hamiltonian,
\begin{equation}
  \label{eq:H_BdG}
  \mathcal{H}(\bm{k}) =
  \begin{pmatrix}
    h(\bm{k})      & \Delta        \\
    \Delta^\dagger & -h^T(-\bm{k}) \\
  \end{pmatrix} ,
\end{equation}
and the Nambu spinors
$\Psi_{\bm{k}} = (c_{\bm{k}}, c_{-\bm{k}}^\dagger)^T$ with
$c_{\bm{k}} = (c_{\bm{k},3/2},\allowbreak c_{\bm{k},1/2},\allowbreak
c_{\bm{k},-1/2},\allowbreak c_{\bm{k},-3/2})^T$, where
$c_{\bm{k},\sigma}$ is the annihilation operator for a fermion with
momentum $\bm{k}$ and spin~$\sigma$.

In this paper, we focus on pairing states in the $T_{2g}$ irrep, where
a general pairing state can be written as
\begin{equation}
  \Delta = \Delta_0\,
  ( l_{yz} \eta_{yz}
  + l_{xz} \eta_{xz}
  + l_{xy} \eta_{xy}) ,
\end{equation}
with the amplitude $\Delta_0$, the three-component order parameter
$\bm{l} = (l_{yz},\allowbreak l_{xz},\allowbreak l_{xy})$, and the gap
matrices
$\eta_{\alpha\beta} = (J_\alpha J_\beta + J_\beta J_\alpha) U_T /
\sqrt{3}$, where
\begin{equation}
  U_T =
  \begin{pmatrix}
    0  & 0 & 0  & 1 \\
    0  & 0 & -1 & 0 \\
    0  & 1 & 0  & 0 \\
    -1 & 0 & 0  & 0 \\
  \end{pmatrix}
\end{equation}
is the unitary part of the time-reversal operator.  From the
fourth-order expansion of the corresponding Landau free energy, four
possible ground states are known: $\bm{l} = (1,0,0)$, $(1,1,1)$,
$(1,\ii,0)$, and $(1,\omega,\omega^2)$ with $\omega = \ee^{2\pi\ii/3}$
(as well as symmetry-related vectors) \cite{Sigrist1991}.  The states
$(1,0,0)$ and $(1,1,1)$ are time-reversal symmetric, whereas the
chiral state $(1,\ii,0)$ and the cyclic state $(1,\omega,\omega^2)$
break TRS and, therefore, support BFSs~\cite{Brydon2018b}.  In the
following, however, we focus on the submanifold of $T_{2g}$ states
spanned by the $\bm{l} = (1,0,0)$ and $(1,\ii,0)$ states by adopting
the mean-field ansatz:
\begin{equation}
  \Delta = \Delta_{yz} \eta_{yz} + \ii \Delta_{xz} \eta_{xz} 
  , \label{eq:meanfield}
\end{equation}
with two \emph{real} variational parameters $\Delta_{xz}$ and
$\Delta_{xz}$. If one of the parameters is zero, we obtain the
TRS-preserving $\bm{l} = (1,0,0)$ state. On the other hand, a TRSB
state is realized if both parameters are nonzero; in particular, the
case of $\Delta_{yz}=\Delta_{xz}$ corresponds to $\bm{l} = (1,\ii,0)$.
Although this restricted ansatz is artificial for a cubic system, we
are motivated by the observation that the $\eta_{xz}$ and $\eta_{yz}$
pairing potentials are the only \textit{s}-wave quintet states in our
cubic model which are also degenerate in hexagonal and tetragonal
crystals.  For example, a chiral \textit{d}-wave state with the same
symmetry is believed to be realized in tetragonal
URu$_2$Si$_2$~\cite{Kasahara2007}. We, therefore, expect our conclusions
to be applicable to any TRSB superconductor with two degenerate
pairing potentials.  The $(1,\ii,0)$ state has an (inflated)
equatorial line node, which should lead to a higher free energy
compared to a state with only (inflated) point nodes.  By considering
the likely less favorable pairing state, we, at worst, underestimate
the stability of the BFSs.  In fact, performing the same analysis for
the pair of $E_g$ states does not result in qualitative changes in the
phase diagram.  The pairing state in the spherically symmetric limit
has been considered in Refs.~\cite{Boettcher2016, Boettcher2018a,
  Herbut2019}.

\subsection{Free energy}

In a weak-coupling approach, the leading pairing instability can be
obtained by direct minimization of the Helmholtz free energy with
respect to the mean fields.  From the BdG Hamiltonian \eqref{eq:H_MF},
we obtain the Helmholtz free energy
\begin{equation}
\label{eq:Helmholtz}
  F = \sum_{\bm{k}} \frac{\tr[\Delta \Delta^\dagger]}{V_0}
  - 2 k_B T \sum_{\bm{k},\nu} \ln\biggl[ 2 \cosh\biggl(\frac{E_{\bm{k},\nu}}
    {2 k_B T} \biggr) \biggr] ,
\end{equation}
where $V_0$ is the attractive pairing interaction in the $T_{2g}$
channel and $E_{\bm{k},\nu}$ are the positive eigenvalues of
$\mathcal{H}(\bm{k})$ in Eq.~\eqref{eq:H_BdG}.  Inserting the
mean-field ansatz from Eq.~\eqref{eq:meanfield}, we numerically
minimize the Helmholtz free energy to obtain the self-consistent
values of $\Delta_{xz}$ and $\Delta_{yz}$.

To compare with our numerical calculation and previous
results~\cite{Agterberg2017}, we also use the complementary approach
of expanding the free energy in the pairing potential to obtain the
Ginzburg-Landau (GL) free energy~\cite{Dahl2007},
\begin{subequations}
\label{eq:GL}
\begin{equation}
  \label{eq:GL-F}
  F = \sum_{\bm{k}} \frac{\tr[\Delta \Delta^\dagger]}{V_0}
  + k_B T \sum_{\bm{k},\,\ii \omega_n}
    \sum_{l=1}^\infty \frac{1}{l} \tr[ ( G_0 \Sigma )^l ] ,
\end{equation}
with
\begin{equation}
  \label{eq:G0-Sigma}
  G_0 =
  \begin{pmatrix}
    G(\bm{k},\ii\omega_n) & 0 \\
    0 & \tilde{G}(\bm{k},\ii\omega_n) \\
  \end{pmatrix}
  \quad\text{and}\quad
  \Sigma =
  \begin{pmatrix}
    0 & \Delta \\
    \Delta^\dagger & 0 \\
  \end{pmatrix}
  ,
\end{equation}
\end{subequations}
where $G$ and $\tilde{G}$ are the particlelike and holelike Green's
functions of the normal-state Hamiltonian $h(\bm{k})$ and
$\ii \omega_n = \ii (2n+1)\pi k_B T$ are the fermionic Matsubara
frequencies.  For this choice of $\Sigma$, all terms with odd $l$
vanish.  The GL free energy can be evaluated analytically, see
Appendix~\ref{sec:GL} for an example calculation and the necessary
approximations.

\subsection{Known limits}

Previous work has revealed the behavior of the model in the limiting
cases of vanishing and strong SOC \cite{Ho1999, Brydon2016,
  Agterberg2017, Brydon2018b, Venderbos2018}. We summarize the results
in the following.

\subsubsection{Vanishing spin-orbit coupling}

The case of vanishing SOC was studied by \citet{Ho1999} in the context
of pairing in fermionic cold atomic gases.  They found that for
\textit{s}-wave quintet pairing, a TRS-preserving state is
energetically favored compared to a TRSB state. To understand this
limit, we first note that the vanishing SOC implies that the
eigenvalues of the normal-state Hamiltonian are fourfold
degenerate. As such, the pairing potential and the normal-state
Hamiltonian can be simultaneously diagonalized by a
momentum-independent spin rotation.  The resulting eigenvalues are
identical to the case of a \textit{s}-wave singlet gap and so the gap
is uniform across the Fermi surface.  For TRSB pairing states, two of
the diagonal entries of the diagonalized pairing potential vanish,
indicating that two of the four degenerate Fermi surfaces remain
ungapped in the superconducting state. On the other hand, a
TRS-preserving state opens a gap on all the Fermi surfaces and is,
thus, energetically favorable.  In real materials, a nonzero SOC is
always present, which lifts the fourfold degeneracy.  We, nevertheless,
expect that for sufficiently weak SOC, the time-reversal-symmetric
state proposed by \citet{Ho1999} persists.

\subsubsection{Strong spin-orbit coupling}
\label{sss:largeSOC}

In the limit where the SOC-induced splitting of the bands is much
larger than the pairing potential, an effective single-band model can
be used for the states close to the Fermi
energy~\cite{Brydon2016}. Specifically, we write the effective BdG
Hamiltonians for the two bands labeled by $\pm$ in the pseudospin
basis as
\begin{equation} \label{eq:Heff}
  \mathcal{H}_{\text{eff},\pm}(\bm{k}) = \begin{pmatrix}
    \epsilon_{\bm{k},\pm}s_0 + \delta H_{\bm{k},\pm} & \pm
    \psi^{\text{intra}}_{\bm{k}}is_y \\
    \mp \psi^{\text{intra}\,\ast}_{\bm{k}}is_y & -\epsilon_{\bm
      k,\pm}s_0 - \delta H^T_{-\bm{k}, \pm}
    \end{pmatrix} ,
\end{equation}
where $s_\mu$ are the Pauli matrices in the pseudospin space. The
effective Hamiltonian describes intraband pseudospin-singlet pairing
with potential,
\begin{equation}
  \label{eq:psieff}
  \psi_{\bm{k}}^{\text{intra}} = \frac{\sqrt{3}\gamma}{2}\,
  \frac{\Delta_{yz} k_y k_z + i\Delta_{xz} k_x k_z}{\sqrt{\sum_i \bigl[ \beta^2 k_i^4
      + (3\gamma^2 - \beta^2) k_i^2 k_{i+1}^2 \bigr]}} .
\end{equation}
The interplay of the quintet pairing with the normal-state spin-orbit
texture gives the intraband potential a \textit{d}-wave form factor,
reflecting the $J=2$ total angular momentum of the Cooper pairs and
imposes a sign difference between the bands. The nodal structure
of the intraband potential favors a TRSB combination of the quintet
states as this gaps out nonintersecting line nodes thereby
enhancing the average gap magnitude and, thus, lowering the free
energy~\cite{Sigrist1991}.  Since the $\eta_{yz}$ pairing potential
leads to line nodes on the $k_y=0$ and $k_z=0$ planes, whereas the
$\eta_{xz}$ state has line nodes on the $k_x=0$ and $k_z=0$ planes,
the $\bm{l} = (1,i,0)$ state is characterized by point nodes along the
$k_z$ axis and a line node on the $k_z=0$ plane.

The diagonal blocks of the effective BdG Hamiltonian in
Eq.~\eqref{eq:Heff} obtain a correction term $\delta H_{\bm{k},\pm}$
from including the effect of interband pairing to second order in
perturbation theory~\cite{Agterberg2017, Brydon2018b,
  Venderbos2018}. This correction has the general form
\begin{equation}
\delta H_{\bm{k},\pm}=  \gamma_{\bm{k},\pm} s_0 + \bm{h}_{\bm{k},\pm}
  \cdot \bm{s} ,
\end{equation}
where $\gamma_{\bm{k},\pm}$ renormalizes the band dispersion and is
always nonzero in the presence of interband pairing, whereas
$\bm{h}_{\bm{k},\pm}$ describes an effective pseudomagnetic field that
is only present for TRSB states. The two contributions can be written
as
\begin{align}
    \gamma_{\bm{k},\pm} &= \frac{1}{2 (\epsilon_{\bm{k},+} - \epsilon_{\bm{k},-})}\,
  \tr[\mathcal{P}_{\bm{k},\pm} \Delta\Delta^\dagger \mathcal{P}_{\bm{k},\pm}] ,\\
  \label{eq:pseudomagn}
  \bm{h}_{\bm{k},\pm} &= \frac{1}{2 (\epsilon_{\bm{k},+} - \epsilon_{\bm{k},-})}\,
  \tr[\bm{s} \mathcal{P}_{\bm{k},\pm} \Delta\Delta^\dagger \mathcal{P}_{\bm{k},\pm}] ,
\end{align}
where $\mathcal{P}_{\bm{k},\pm}$ are projection operators on the
normal-state Hilbert spaces of the $\pm$ bands.  The pseudomagnetic
field is crucial for the appearance of BFSs as can be seen from the
dispersion in the effective low-energy model,
\begin{equation}
  \label{eq:Eeff}
  E_{a,b,\pm} = a \abs{\bm{h}_{\bm{k},\pm}}
    + b \sqrt{[\epsilon_{\bm{k},\pm} + \gamma_{\bm{k},\pm}]^2
    + \abs{\psi^{\text{intra}}_{\bm{k}}}^2} ,
\end{equation}
where $a$ and $b$ are independently chosen to be $\pm 1$, giving four
bands. In the absence of the pseudomagnetic field, a node occurs where
the square root vanishes, but the pseudomagnetic field is generally
nonzero at these momenta. This lifts the pseudospin degeneracy by
shifting the pseudospin-up and pseudospin-down bands in opposite
directions and leads to the formation of BFSs \cite{Agterberg2017,
  Brydon2018b}.  Although this increases the free energy of the TRSB
state, for sufficiently small $\abs{\bm{h}_{\bm{k},\pm}}$, it should
not cause a transition to a TRS-preserving phase, since the energy
difference between the lowest TRSB and TRS-preserving states is
generically finite. In particular, from Eq.~\eqref{eq:pseudomagn} we
expect that a TRSB state with BFSs is stable for
$\abs{\Delta_{yz}}, \abs{\Delta_{xz}} \ll
\abs{\epsilon_{\bm{k},+}-\epsilon_{\bm{k},-}}$.

\section{Phase diagram}
\label{sec:trsb}

\begin{figure}[tb]
  \centering
  \includegraphics{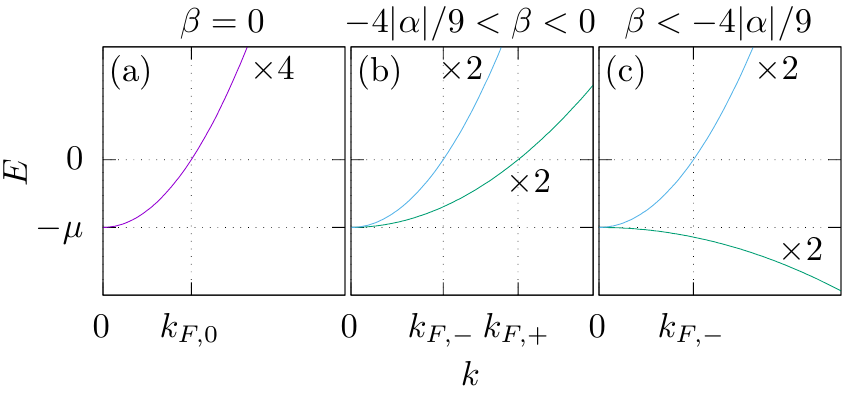}
  \caption{Representative dispersion relations for the spherically
    symmetric model described by Eq.~\eqref{eq:hspherical}. (a)
    Without SOC, the bands are fourfold degenerate, and there is a
    single Fermi surface with wave vector
    $k_{F,0}=\sqrt{\mu/\alpha}$. Note that we take $\alpha>0$ so that
    the band has positive effective mass.  (b) Turning on the SOC
    lifts the fourfold degeneracy, yielding two doubly degenerate
    quadratically dispersing bands with positive effective mass. There
    are now two Fermi surfaces with wave vectors
    $k_{F,\pm} = \sqrt{\mu/(\alpha + 5\beta/4 \pm \beta)}$. (c) For
    $\beta < -4\alpha/9$, the effective mass of one of the bands
    becomes negative, and there is only a single Fermi surface.}
  \label{fig:cartoon}
\end{figure}

We start by considering the case of spherically symmetric SOC, i.e.,
$\beta = \gamma$, and later generalize to the case of cubic
anisotropy.  In the spherical limit, the normal-state Hamiltonian
simplifies to
\begin{equation}
  h(\bm{k}) = (\alpha \abs{\bm{k}}^2 - \mu)\, \openone_4 + \beta (\bm{k}
  \cdot \bm{J})^2 . \label{eq:hspherical}
\end{equation}
Representative examples of the normal-state band structure are shown
in Fig.~\ref{fig:cartoon}.

\begin{figure*}[tb]
  \centering
  \includegraphics{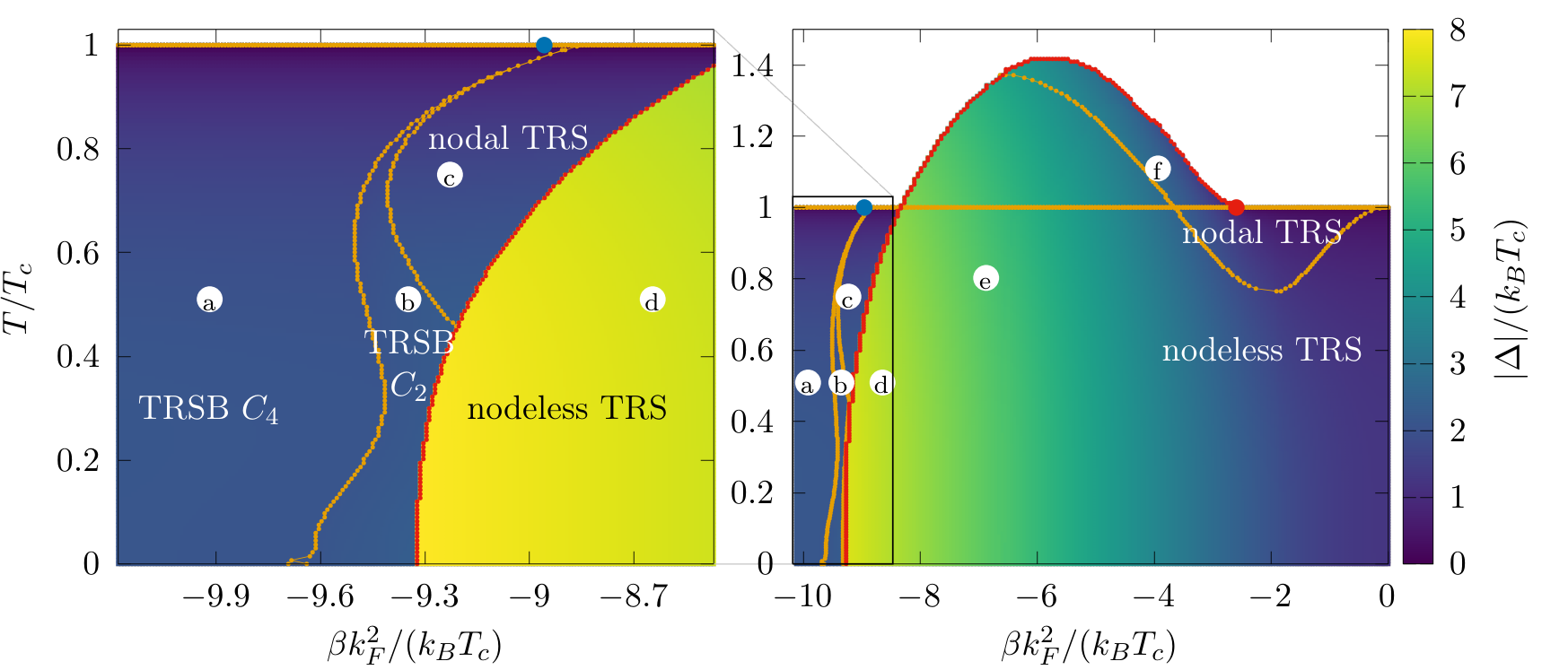}
  \caption{Phase diagram for the $T_{2g}$ pairing states given by
    Eq.~\eqref{eq:meanfield} on the SOC-temperature plane.  The color
    code indicates the gap magnitude
    $\sqrt{\Delta_{xz}^2+\Delta_{yz}^2}$ where brighter colors mean
    larger gaps and white means no superconductivity.  The horizontal
    line at $T/T_c = 1$ denotes the critical temperature $T_c$
    predicted by GL theory.  Lines of first-order (second-order) phase
    transitions are indicated in red (orange).  The blue dot in both
    panels indicates the point of TRSB from GL theory, the red dot in
    the panel on the right denotes the onset of the first-order phase transition
    estimated by GL theory. The left panel is a zoom of the box in the
    right panel.  The SOC strength $\beta$ is plotted as an effective
    spin-orbit energy $\beta k_F^2/(k_B T_c)$ where
    $k_F^2 = \mu/(\alpha + 5\beta/4)$.}
  \label{fig:spher}
\end{figure*}

\begin{figure*}[tb]
  \includegraphics{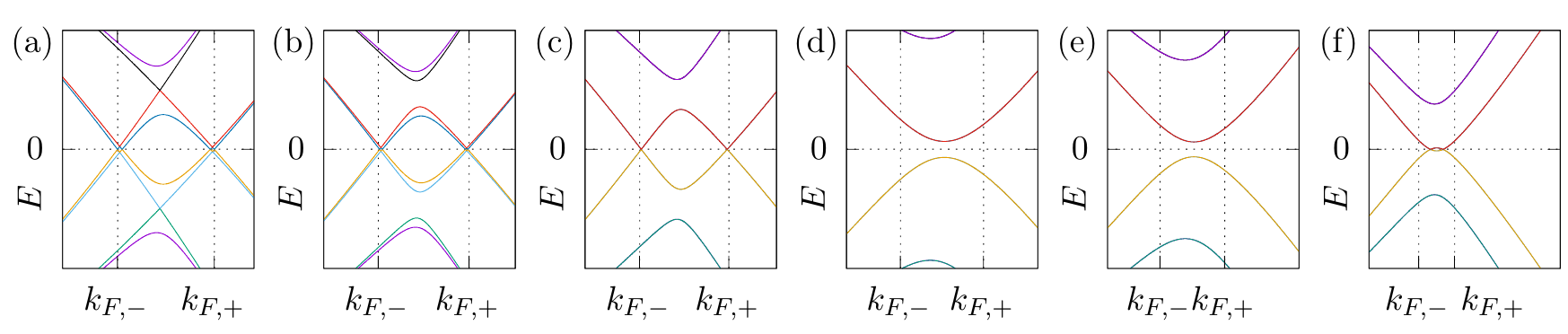}
  \caption{(a)--(f) Band structure in the vicinity of the Fermi energy
    for parameter sets indicated by the corresponding labels in
    Fig.~\ref{fig:spher} along the $[100]$ direction where we expect
    nodes in a nodal state.  The Fermi wave vectors are given by
    $k_{F,\pm} = \sqrt{\mu/(\alpha + 5\beta/4 \pm \beta)}$.}
  \label{fig:insets}
\end{figure*}

In Fig.~\ref{fig:spher}, we present the phase diagram as a function of
temperature and SOC strength.  Figures~\ref{fig:insets}(a)--\ref{fig:insets}(f) show
the band structure around the Fermi surface in the $[100]$ direction
where we anticipate the appearance of nodes from the projected gap in
Eq.~\eqref{eq:psieff}. Any gaps in the spectrum along this direction
at nonzero SOC strength are, therefore, due entirely to the interband
pairing potential.  To obtain comparable results over a wide range of
SOC strengths, we fix the critical temperature $T_c$ and vary the
attractive interaction $V_0$ such that the second-order coefficient of
the GL free energy vanishes at the chosen $T_c$. This eliminates
effects due to the changing DOS at the Fermi energy as the SOC is
varied.

Starting at $\beta = 0$, we find the fully gapped TRS-preserving state
(``nodeless TRS'') predicted by \citet{Ho1999}.  Switching on the SOC,
we observe that the gap just below the critical temperature has nodes
(``nodal TRS''), but the nodeless TRS state is recovered at lower
temperatures. The nodal behavior arises as the SOC lifts the fourfold
degeneracy of the bands, making a distinction between inter- and
intraband pairings possible. Close to the critical temperature, the
strength of the pairing potential is much smaller than the band
splitting so that the gap at the Fermi surface is controlled by the
nodal intraband pairing potential in Eq.~\eqref{eq:psieff}.  However,
as the pairing potential grows upon lowering the temperature, the
interband potential shifts the nodes away from the Fermi surfaces at
$k_{F,\pm} = \sqrt{\mu/(\alpha + 5\beta/4 \pm \beta)}$ as seen in the
band structure at point~(f) in Fig.~\ref{fig:insets}.  At a critical
value of the pairing potential, the nodes meet and annihilate, marking
the recovery of the nodeless TRS phase.

A further increase in SOC leads to an enhancement of the critical
temperature over the one anticipated from the second-order coefficient
of the GL free energy, implying a first-order transition between the
normal and the superconducting states.  The presence of the first-order
phase transition is confirmed by computing the position of the
tricritical point from GL theory, i.e., the point where the
fourth-order coefficient turns negative.  We find very good agreement
between the numerical calculation and our GL theory (cf.\ the red dot
in the right panel of Fig.~\ref{fig:spher}). In this region, the
magnitude of the pairing potential
$\sqrt{\Delta_{xz}^2+\Delta_{yz}^2}$ is much larger than expected from
BCS theory, and the very large interband pairing potentials ensure a
full gap as shown by points~(d) and~(e) in Fig.~\ref{fig:spher}. We,
hence, refer to this state as the ``large-gap'' phase, in contrast to
the other, ``small-gap'' phases. The origin of the first-order phase
transition is discussed in Sec.~\ref{sec:toy-model}.

\begin{figure}
  \centering
  \includegraphics{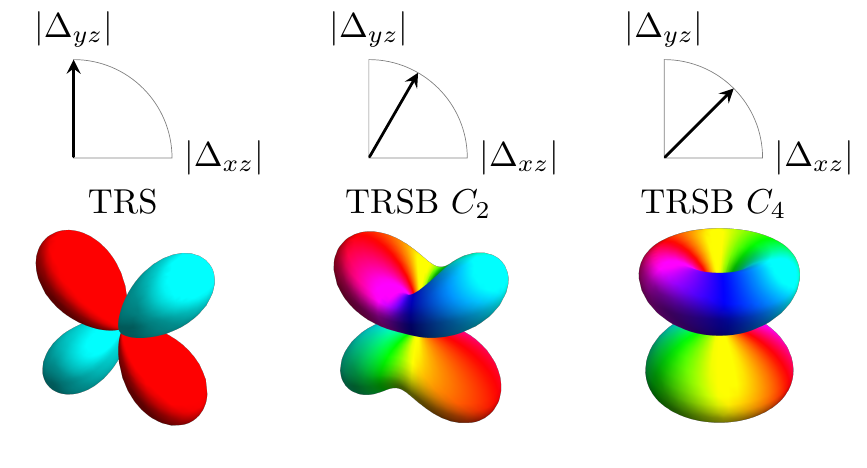}
  \caption{Sketch of the pairing amplitudes $\abs{\Delta_{xz}}$ and
    $\abs{\Delta_{yz}}$ and of the gap structure in the nodal TRS, TRSB
    $C_2$, and TRSB $C_4$ phases, see Fig.~\ref{fig:spher}. The TRSB
    $C_2$ state breaks both TRS and $C_4$ symmetry.}
  \label{fig:phase-cartoon}
\end{figure}

Upon increasing the SOC strength beyond
$\beta k_F^2 \approx -8.4 k_B T_c$, there is an abrupt drop in the
magnitude of the gap, and the nodeless TRS phase gives way to a nodal
state.  Close to $T_c$, this state marked by ``nodal TRS'' has a gap
that is well approximated by Eq.~\eqref{eq:psieff} and exhibits line
nodes [point~(e) in Figs.~\ref{fig:spher}
and~\ref{fig:insets}]. Further below $T_c$, we enter a phase which
breaks TRS but where the two gap parameters $\Delta_{yz}$ and
$\Delta_{xz}$ have unequal magnitude.  We label this the ``TRSB
$C_2$'' state because the unequal gap magnitudes yield a spectrum with
only $C_2$ rotational symmetry about the $z$ axis.  The magnitudes of
$\Delta_{yz}$ and $\Delta_{xz}$ converge as the SOC is increased, thus,
realizing the ``TRSB $C_4$'' state where the spectrum has $C_4$
rotational symmetry about the $z$ axis.  This is the
$\bm{l} = (1,i,0)$ state and is consistent with predictions of the
strong-SOC limit discussed in Sec.~\ref{sss:largeSOC}.  The
intermediate TRSB $C_2$ state can be visualized as a continuous
rotation of the vector $\bm{l}$ from $(1,0,0)$ to $(1,i,0)$, see
Fig.~\ref{fig:phase-cartoon}.  The boundary of the TRSB $C_4$
phase shows reentrant behavior, but it is realized at all temperatures
for $\beta k_F^2 \approx -9.7 k_B T_c$.  Both the TRSB $C_4$ and the
$C_2$ phases display BFSs.

The critical value of the SOC strength for which the TRSB state
becomes stable just below $T_c$ is estimated from an expansion of the
GL free energy to fourth order at
$\beta k_F^2 \approx -8.957 k_B T_c$.  This estimate is shown as the
blue dot in both panels of Fig.~\ref{fig:spher} and is in excellent
agreement with the mean-field theory.  A previous
analysis~\cite{Agterberg2017} had estimated this critical strength to
be $\beta k_F^2 \approx -11.572 k_B T_c$ (expressed in our units) and,
therefore, overestimated it by about $30\%$. The disagreement stems
from the approximate treatment of the band splitting in
Ref.~\cite{Agterberg2017}. Nevertheless, we confirm that the TRSB
state is realized at moderate values of the SOC strength.

\subsection{Effects of cubic anisotropy}
\label{sec:cubic}

\begin{figure}[tb]
  \includegraphics{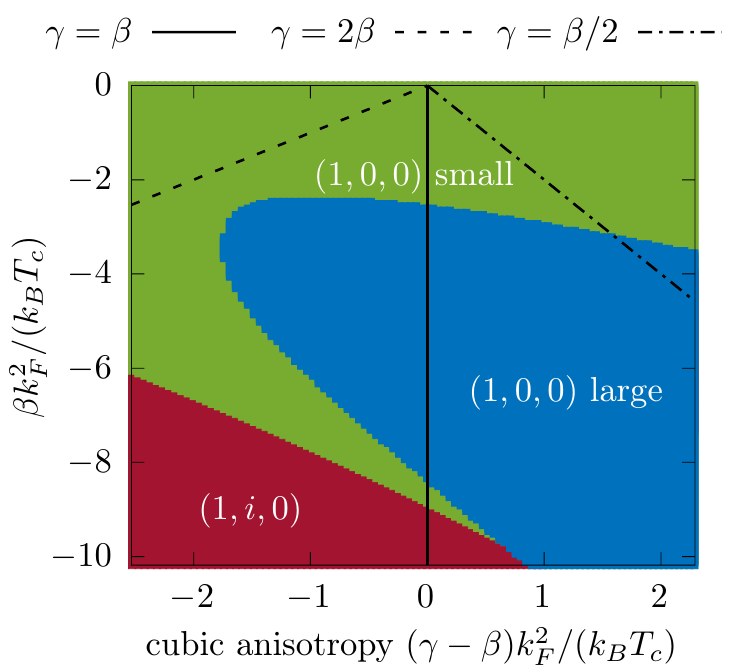}
  \caption{Phase diagram just below the critical temperature as a
    function of SOC strength and cubic anisotropy.  ``$(1,0,0)$
    small'' and ``$(1,0,0)$ large'' refer to the TRS-preserving phase,
    whereas ``$(1,i,0)$'' specifies the TRSB $C_4$ phase with BFSs.}
  \label{fig:F4}
\end{figure}

Cubic anisotropy is introduced in our model by setting
$\gamma \neq \beta$ in the Luttinger-Kohn Hamiltonian. In
Fig.~\ref{fig:F4}, we show the pairing state realized just below the
critical temperature as a function of $\beta$ and the cubic anisotropy
parameter $\gamma-\beta$. Note that the transition into the large-gap
phase is of first order and the critical temperature, therefore,
exceeds the temperature at which the second-order coefficient in the
GL expansion changes sign. As can be seen, there is a pronounced
asymmetry between the cases of $\abs{\gamma}>\abs{\beta}$ and
$\abs{\gamma}<\abs{\beta}$: The region of first-order transitions into
the large-gap phase is suppressed for $\abs{\gamma}>\abs{\beta}$ and
disappears entirely for sufficiently strong $\gamma$, and the TRSB
state occurs at smaller values of the SOC strength
$\abs{\beta}$. These trends are reversed for $\gamma>\beta$.

\begin{figure}[t!b]
  \includegraphics{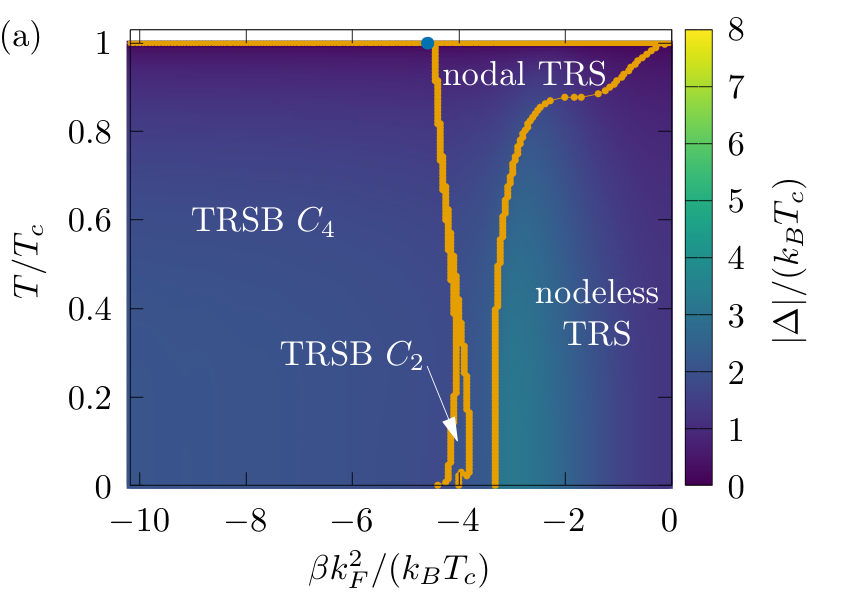}\\[2ex]
  \includegraphics{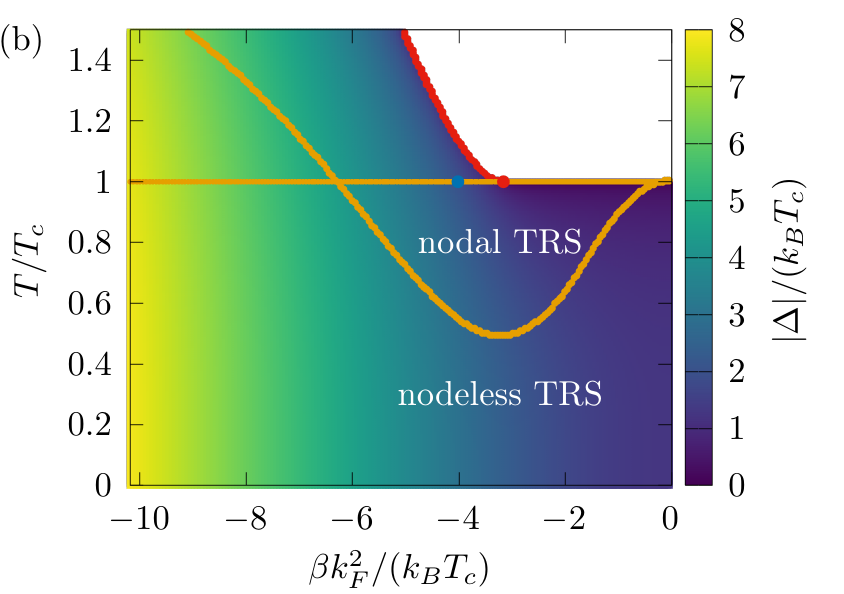}
  \caption{Phase diagrams on the SOC-temperature plane along two
    lines (a) $\gamma = 2\beta$ and (b) $\gamma = \beta/2$.  The
    color code represents the gap magnitude where brighter colors
    mean larger gaps and white means no superconductivity.  Lines of
    first-order (second-order) phase transitions are indicated in red
    (orange).  For $\gamma=2\beta$, panel (a), there is no first-order
    phase transition at $T_c$.  Below $T_c$, we find a large-gap phase,
    but the transition to it is not of first order.  For
    $\gamma = \beta/2$, panel (b), the large-gap phase occurs.  The
    critical temperature is strongly enhanced, and larger gaps are
    found.  In both panels, the blue dot at $T_c$ is the point of TRSB,
    and the red dot at $T_c$ is the tricritical point as predicted by
    the GL free energy.}
  \label{fig:cubic}
\end{figure}

In Fig.~\ref{fig:cubic}, we show temperature-dependent phase diagrams
along two lines $\gamma = 2\beta$ and $\gamma = \beta/2$.  Along the
cut $\gamma = 2\beta$, there is no first-order phase transition.  The
change in gap magnitude along the nodeless to nodal transition is
steep but not abrupt.  This transition is accompanied by the
disappearance of nodes.  The intermediate $C_2$ phase is also heavily
suppressed.  Along the other cut $\gamma = \beta/2$, we do not recover
the small-gap phase within the boundaries of the graph.  Therefore, we
also do not observe the point of TRSB as predicted by GL theory.

The phase diagram in Fig.~\ref{fig:F4} can be understood by looking at
the expression for the effective intraband pairing
Eq.~\eqref{eq:psieff}.  We find that the magnitude of the intraband
pairing is proportional to $\gamma$, i.e., larger (smaller) $\gamma$
means stronger (weaker) intraband pairing compared to interband
pairing. The existence of the large-gap phase depends on the ratio
between intra- and interband pairings as we will discuss in the next
section.

\subsection{Origin of the first-order transition}
\label{sec:toy-model}

The first-order phase transition into the large-gap phase shown in
Figs.~\ref{fig:spher}, \ref{fig:F4}, and \ref{fig:cubic} is one of the
most surprising features of the phase diagram of our model. The
inclusion of cubic anisotropy reveals that it is not generic, however,
but rather depends upon the balance between the two spin-orbit terms.
In this section, we show that the first-order transition is controlled
by the relative strengths of the intra- and interband pairing
potentials, which, in turn, depends on the SOC strengths as noted
above.

The first-order transition can be understood based on a simplified
model with two bands in which we fix the ratio of the inter- and
intraband pairing potentials. In this model, the normal-state bands
have the dispersions
$\xi_{\bm{k},\pm} = (1\pm \delta)\epsilon_{\bm{k}} - \mu$, where
$\delta$ parametrizes the band splitting and the precise form of
$\epsilon_{\bm{k}}$ is unimportant. The splitting parameter $\delta$
plays a role analogous to the SOC strength in the full model where
the band splitting is characterized by differing effective masses of
the Luttinger-Kohn bands as illustrated in Fig.~\ref{fig:cartoon}.

Since the interband and intraband pairing potentials are obtained by
projecting $\Delta$ from Eq.~\eqref{eq:meanfield} into the band basis,
the relative strength of the interband and intraband pairing
potentials is determined by details of the normal-state band
structure.  To represent this aspect, we write the pairing potential
in the band basis as
\begin{equation}
  \label{eq:psitoy}
  \Delta = \eta
  \begin{pmatrix}
    r            & \sqrt{1-r^2} \\
    \sqrt{1-r^2} & -r           \\
  \end{pmatrix},
\end{equation}
where $\eta$ is the magnitude of the pairing potential, and the
coefficient $r$ controls the relative strength of intra- and interband
pairings: $r = 0$ corresponds to pure interband pairing and $r = 1$ to
pure intraband pairing. The intraband pairing has opposite signs in
each band, in agreement with Eq.~\eqref{eq:psieff}. Since the
first-order transition only occurs into a TRS-preserving state, in the
following we assume that $r$ and $\eta$ are real.  Note that for the
pairing potential in Eq.~\eqref{eq:psitoy}, the ratio between intra-
and interband pairings is momentum independent. In contrast, in the
full model, this quantity varies across the Fermi surface.  We can
nevertheless define this ratio for the full model in terms of the
Fermi-surface average,
\begin{equation} \label{eq:avr}
  r^2 = \frac{1}{\Delta_{xz}^2+\Delta_{yz}^2}
  \int \frac{\diff\Omega}{4\pi}\:\abs{\psi_{\bm{k}}^{\text{intra}}}^2 .
\end{equation}

\begin{figure}[tb]
  \includegraphics{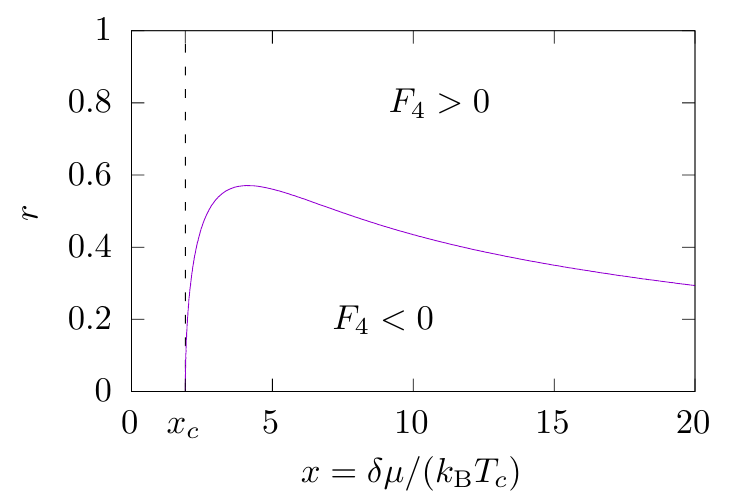}
  \caption{Phase diagram of the simple model as a function of the
    pairing ratio $r$ and the band splitting $\delta$.  In this model,
    first-order phase transitions are only possible for $x > x_c$
    because only then there is a region where $F_4 < 0$.}
  \label{fig:toy}
\end{figure}

The GL expansion of the free energy of the simple model gives a Taylor
series in the parameter $\eta$,
\begin{equation} \label{eq:GLexpand}
  F = F_2\eta^2 + F_4\eta^4 + \mathcal{O}(\eta^6) ,
\end{equation}
where expressions for the coefficients $F_2$ and $F_4$ can be obtained
from Eqs.~\eqref{eq:GL-F} and~\eqref{eq:G0-Sigma}. A negative sign of
the fourth-order coefficient $F_4$ indicates that the transition into
the superconducting state is of first order. We show the variation of
the sign of this coefficient as a function of the parameters $r$ and
$\delta$ in Fig.~\ref{fig:toy}. For sufficiently small intraband
pairing strength $r$, we find that $F_4$ is positive at small band
splitting $\delta$, becomes negative for increasing $\delta$, and
finally returns to a positive value. Assuming that higher-order terms
in the GL expansion can be ignored, this indicates that the phase
transition becomes discontinuous beyond a critical band splitting, but
a continuous transition is recovered as the band splitting is further
increased.

The conclusions for the simple model are broadly in agreement with the
phase diagrams for the full model in Figs.~\ref{fig:spher}
and~\ref{fig:F4}.  Equation \eqref{eq:avr} gives $r=1/\sqrt{5}$ for
the full model in the spherical limit.  According to the simple model,
the phase transition at this value of $r$ becomes discontinuous at
$\abs{x} = \abs{\delta\mu/(k_B T_c)} \approx 2.460$, which is in very
good agreement with the location of the tricritcal point for the full
model at $\abs{x} \approx 2.594$ (red dot in Fig.~\ref{fig:spher})
where the effective band splitting is given by
$\delta = \beta/(\alpha + 5\beta/4)$. The simple model also explains
the asymmetric effect of the cubic anisotropy seen in
Fig.~\ref{fig:F4}: For $\abs{\gamma}>\abs{\beta}$, the intraband
pairing potential is enhanced, which, in turn, increases the value of
$r$ and, thus, suppresses the first-order transition. Conversely,
$\abs{\gamma}<\abs{\beta}$ reduces the intraband pairing potential and,
thus, $r$ and favors the first-order transition.

The simple model and our full results agree in showing that a
second-order transition is recovered at sufficiently large values of
the band splitting $\delta$. The reappearance of the second-order
transition in the full model, however, does not occur with a
tricritical point, but rather with a discontinuous jump in the minimum
of the free energy from a large value of the gap magnitude
$(\Delta_{xz}^2 + \Delta_{yz}^2)^{1/2}$ (large-gap phase) to a minimum
at a small value of the gap magnitude (small-gap phase).  Properly
capturing this behavior in the simple model would require extending
the GL expansion in Eq.~\eqref{eq:GLexpand} to, at least, eighth order
in~$\eta$.

\section{Properties of the time-reversal-symmetry-breaking state
\label{sec:trsbstate}}

We now investigate features of the TRSB $C_4$ state. We choose the
parameter set labeled by (c) in Figs.~\ref{fig:spher}
and~\ref{fig:insets}.  In this case, we can set
$\Delta_{xz}=\Delta_{yz}=\Delta_0$, and so, the pairing potential is
$\Delta = \Delta_0\, (\eta_{yz} + \ii \eta_{xz})$.

\begin{figure}[tb]
  \includegraphics{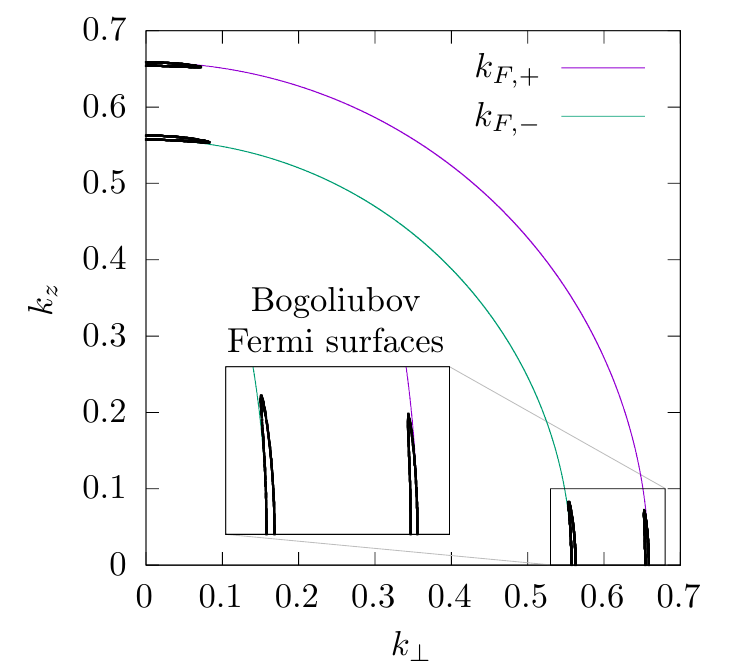}
  \caption{BFSs for the parameters labeled by (c) in
    Figs.~\ref{fig:spher} and~\ref{fig:insets} (heavy black lines).
    The colored lines denote the normal-state Fermi surfaces of the
    $+$ and $-$ bands. $k_\perp$ is the radial component of the
    momentum on the $k_x k_y$-plane.}
  \label{fig:fs}
\end{figure}

\subsection{Bogoliubov Fermi surfaces}

First, we map out the BFSs by searching for vanishing energy
eigenvalues.  Thanks to rotational symmetry around the $z$ axis and
inversion symmetry, we can restrict ourselves to the first octant. The
resulting nodal surfaces are shown in Fig.~\ref{fig:fs}.  In the TRSB
$C_4$ state, the magnitude of the pseudomagnetic field in
Eq.~\eqref{eq:pseudomagn} is
\begin{equation}
  \abs{\bm{h}(\bm{k})} = \frac{4 \abs{\Delta_0}^2}{(\epsilon_{\bm{k},+}
    - \epsilon_{\bm{k},-})^2}\,
  \beta\, \sqrt{\abs{\bm{k}}^4 - 3 (k_x^2 + k_y^2) k_z^2} .
\end{equation}
The size of the BFSs scales with the magnitude of the pseudomagnetic
field. Since this field is inversely proportional to the band
splitting squared, which grows as $\abs{\bm{k}}^2$, the inner BFS is
larger than the outer one, see the inset of Fig.~\ref{fig:fs}.  The
pseudomagnetic field has the largest magnitude close to the boundary with
the TRSB $C_2$ state since this corresponds to the smallest band
splitting for which the TRSB $C_4$ state is stable.  Here, the BFSs
have the largest volume and are, therefore, clearly distinguishable from
line and point nodes.

\begin{figure}[tb]
  \includegraphics{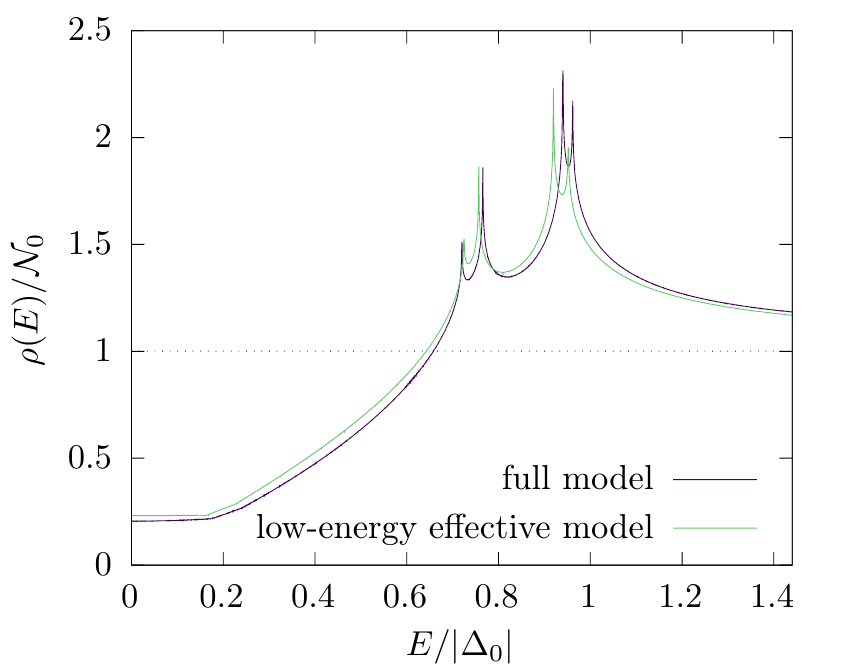}
  \caption{DOS in the superconducting state for the parameters labeled
    by (c) in Figs.~\ref{fig:spher} and~\ref{fig:insets}, based on a
    full two-band calculation (black curve) and on a low-energy
    single-band approximation (green curve). The results of the two
    approaches agree very well.  The residual DOS at zero energy is as
    large as 20\% of the normal-state DOS at the Fermi energy
    $\mathcal{N}_0=\sqrt{\mu}/2(\alpha + 5\beta/4)^{3/2}$.}
  \label{fig:dos}
\end{figure}

The existence of BFSs leads to a nonzero DOS at zero energy, which is
not expected for clean superconductors.  We compute the DOS
numerically from the mean-field dispersion and analytically using
the low-energy dispersion from Eq.~\eqref{eq:Eeff}.  In the absence of
cubic anisotropy, close to the Fermi surface, $\bm{h}(\bm{k})$,
$\gamma_{\bm{k}}$, and $\psi_{\bm{k},\pm}$ only depend upon the polar
angle $\theta$.  The DOS in the $\pm$ band is, thus,
\begin{equation} \label{eq:DOS}
  \rho_\pm(E) = \mathcal{N}_{0,\pm}
    \sum_{a,b} \int_0^{\pi}
  \frac{\bigl| E - a \abs{\bm{h}_\pm(\theta)} \bigr|\, \sin\theta \diff\theta}
    {\sqrt{(E - a \abs{\bm{h}_\pm(\theta)})^2 - \abs{\psi_\pm(\theta)}^2}} ,
\end{equation}
where we have assumed the normal-state DOS $\mathcal{N}_{0,\pm}$ to be
constant in the range of the superconducting gap.  Evaluating
Eq.~\eqref{eq:DOS}, we find excellent agreement with the numerical
results, as shown in Fig.~\ref{fig:dos}. In particular, we clearly see
a large residual DOS at zero energy in the superconducting gap of up
to 20\% of the normal-state DOS. The flat DOS at zero energy results
from the lifting of the pseudospin degeneracy by the pseudomagnetic
field $\bm{h}$. This shifts the DOS for each pseudospin species,
leading to the scaling
$\rho(E) \propto (\abs{E + \abs{\bm{h}}} + \abs{E - \abs{\bm{h}}})/2$ instead
of $\rho(E) \propto \abs{E}$, as would be the case for line nodes.
This gives a constant DOS for $-\abs{\bm{h}} < E < \abs{\bm{h}}$, as
previously reported in Ref.~\cite{Lapp2019}.  The effect of the
pseudomagnetic field is also seen in the splitting of the coherence
peaks: In the absence of the pseudomagnetic field, we expect a single
coherence peak at $\abs{E}=\Delta_0$. Upon adding the pseudomagnetic
field, it is split into four coherence peaks at
$\Delta_0 + \abs{\bm{h}_{\pm}(\theta = \pi/4)}$ and
$\Delta_0 - \abs{\bm{h}_{\pm}(\theta = \pi/4)}$, where $\theta = \pi/4$ is
the angle of maximum gap. Since the pseudomagnetic field has different
magnitude at the two Fermi surfaces, these two peaks are, in turn,
weakly split.

A residual DOS in an unconventional superconductor can also arise due
to the presence of impurities~\cite{Mineev1999}.  We can estimate the
required concentration of impurities to achieve a zero-energy residual
DOS as large as 20\% of the normal-state DOS within the
self-consistent Born approximation.  Using the exact results for the
polar phase of $^3$He, which also has an equatorial line node, we
estimate that the required concentration of impurities would
approximately result in a 40\% suppression of $T_c$ compared to the
clean limit.  It should be possible to rule out the effect of
impurities by considering the residual DOS as a function of $T_c$ for
different samples.  We also emphasize that the splitting of the
coherence peaks seen in Fig.~\ref{fig:dos} cannot be explained by
impurity effects.

\subsection{Induced magnetic order parameter}

\begin{figure}[tb]
  \includegraphics{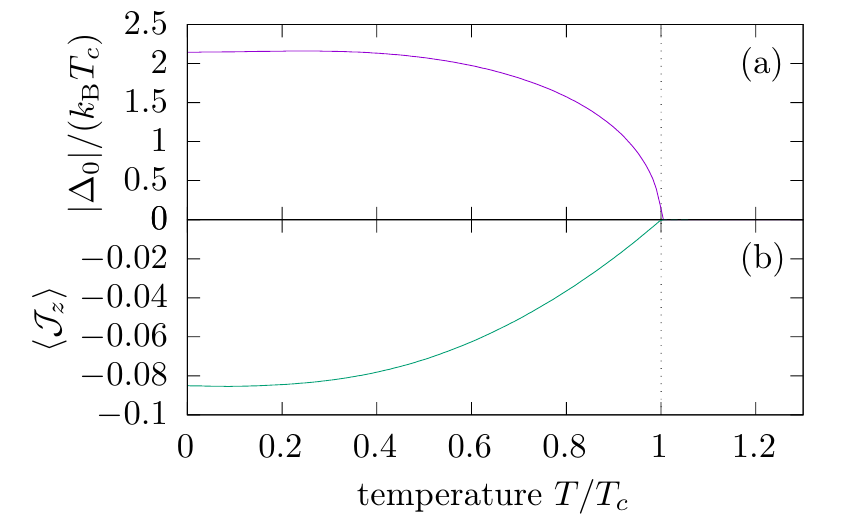}
  \caption{(a) Gap parameter and (b) induced magnetic order parameter
    as functions of temperature in the TRSB phase and with the SOC
    labeled by (c) in Figs.~\ref{fig:spher} and~\ref{fig:insets}.}
  \label{fig:SI}
\end{figure}

As pointed out in Ref.~\cite{Brydon2018b}, the pseudomagnetic field can be
interpreted as manifesting a subdominant secondary magnetic order
parameter, which is induced by the superconductivity. This subdominant
order is related to the time-reversal odd part of the gap product,
\begin{equation}
  \Delta \Delta^\dagger - U_T\Delta^\ast \Delta^TU_T^\dagger =
  \frac{4}{3}\, \Delta_0^2 (7 J_z - 4 J_z^3)
  \mathrel{\equiv} 2\Delta_0^2 \mathcal{J}_z .
\label{eq:subdominant}
\end{equation}
In Fig.~\ref{fig:SI}, we show the expectation value of $\mathcal{J}_z$
together with the superconducting gap as functions of temperature. The
superconductivity and magnetism appear together but their temperature
dependence close to the critical temperature is notably different:
whereas the gap magnitude scales as $\Delta_0 \sim \abs{T-T_c}^{1/2}$, the
expectation value of $\mathcal{J}_z$ scales as
$\<\mathcal{J}_z> \sim \abs{T-T_c}$. This linear temperature dependence
close to $T_c$ reflects its relation to the gap product in
Eq.~\eqref{eq:subdominant}.

The finite expectation value of $\mathcal{J}_z$ generically leads to a
finite pseudomagnetic field in Eq.~\eqref{eq:pseudomagn} and, thus, to a
momentum-dependent spin polarization. To understand the interplay
between magnetism and superconductivity, we include a magnetic order
parameter $m_z$ in the channel that couples to superconductivity in
the GL expansion.  To that end, following~\cite{Dahl2007}, we redefine:
\begin{equation}
\label{eq:Sigma_new}
  \Sigma =
  \begin{pmatrix}
    \mathcal{M}_z & \Delta \\
    \Delta^\dagger & -\mathcal{M}_z^T \\
  \end{pmatrix} ,
\end{equation}
in Eq.\ \eqref{eq:G0-Sigma}, where
$\mathcal{M}_z = m_z \mathcal{J}_z$.  The lowest-order coupling
between the superconducting and magnetic order parameters occurs at
third order and has the form
\begin{equation}
  \ii F_3m_z (\Delta_{xz}\Delta_{yz}^* - \Delta_{xz}^* \Delta_{yz} ),
\end{equation}
which clearly indicates that the TRSB superconducting state induces
the magnetism.  The lengthy expression for the coefficient $F_3$ is
presented in Appendix~\ref{sec:F3}. In particular, we must introduce a
cutoff $\Lambda$ of the attractive pairing interaction to account for
particle-hole asymmetry in the normal state.  In the limit where the
band splitting and cutoff are much larger than $k_BT_c$ (i.e., the
conditions under which the TRSB state is stable), the coefficient
simplifies to
\begin{equation}
  \label{eq:F3}
  F_3 = \frac{\mathcal{N}_0}{\mu}\,
  \frac{48}{5}
  \biggl[1-  \frac{\ln\frac{2\Lambda e^\gamma}{\pi k_BT_c}}
    {3(1-\tilde{\beta}^2)}
  -\frac{1}{4} \ln
    \biggl(1 + \frac{\Lambda^2}{\tilde{\beta}^2\mu^2}\biggr)\biggr]  
    ,
\end{equation}
where $\mathcal{N}_0 = \sqrt{\mu}/2\, (\alpha + 5\beta/4)^{3/2}$ is
the normal-state DOS at the Fermi energy,
$\tilde\beta = \beta/(\alpha + 5\beta/4)$, $\tilde{\beta}\mu$ is the
band splitting, and $\gamma$ is the Euler-Mascheroni constant. To
understand this result, we note that the magnetic order paramater
$m_z$ couples to $\mathcal{J}_z$, which is not diagonal in the band
basis but has both interband and intraband components. The intraband
component directly couples to the pseudomagnetic field generated by
the interband pairing potentials and gives a cutoff-independent
contribution to $F_3$. On the other hand, the interband components of
the magnetic order couple to both the intraband and the interband pairing
potentials and give the cutoff-dependent contribution, see Appendix
\ref{sec:F3} for details.  These two contributions have opposite signs
and the contribution from the interband component is likely dominant
when $\Lambda \gg k_BT_c$.

It is interesting to compare our results to the more familiar case of
coupling between ferromagnetic and superconducting order parameters in
a single-band TRSB superconductor~\cite{Mineev1999}. A similar GL
expansion of the free energy in that case also gives a third-order
coupling term with coefficient proportional to $\mathcal{N}_0/\mu$,
which implies that the magnetization in the superconducting state is
on the order of $\Delta_0^2/\mu^2$ and is, hence, expected to be weak.
This property is thought to be generic for TRSB superconductors
\cite{Cross1975,Mermin1980}. In the present case, it can be understood
as being due to the fact that the $j = 1/2$ and $j = -3/2$
quasiparticles do not participate in the pairing.  The spin of these
unpaired quasiparticles then compensates the polarization of the
Cooper pairs as is the case for a spin-$1/2$ superconductor where
only the up spin is paired and the unpaired down spin compensates the
polarization~\cite{Mineev1999}. The presence of a BFS, therefore, does
not imply a strong magnetization of the superconductor.

\section{Summary and conclusions}
\label{sec:conclusion}

In this paper, we have used BCS mean-field theory to study the
evolution of the quintet superconducting state in the paradigmatic
Luttinger-Kohn model as a function of the SOC strength. We find a rich
phase diagram in the SOC-temperature plane.  For weak SOC, a
time-reversal-symmetric superconducting state is realized. Upon
increasing the SOC strength, the transition into the superconducting
state becomes first order. The origin of the first-order transition is
the competition between inter- and intraband pairings, which is
controlled by the cubic anisotropy of the SOC; for sufficiently
anisotropic SOC, the first-order transition can be completely
suppressed.  Upon further increasing the SOC strength, first a
second-order transition is recovered, and finally a TRSB pairing state
is stabilized.  At low temperatures, the TRSB state displays reentrant
behavior as well as a first-order transition into the TRS-preserving
state.

The TRSB state exhibits BFSs and a residual DOS at the Fermi energy,
which can be as large as 20\% of the normal-state DOS.  The TRSB
pairing state induces a subdominant magnetic order parameter, which we
find to be small even if the residual DOS is sizable, consistent with
the general result that TRSB superconductors have weak intrinsic
magnetization.

Our analysis establishes that a pairing state with BFSs can be
thermodynamically stable, even when the residual DOS at the Fermi
energy due to the BFSs is a sizable fraction of the normal-state
DOS. This result is encouraging for experimental searches for BFSs as
it shows that the residual DOS due to the BFSs can be of detectable
magnitude. Since the size of the BFSs is controlled by the ratio of
the interband pairing potential to the band splitting, materials where
this ratio is as large as possible are the best candidates. This
suggests that heavy-fermion superconductors are promising. It is,
therefore, intriguing that a residual DOS has been observed in
URu$_2$Si$_2$~\cite{Kasahara2007} and UTe$_2$~\cite{Ran2019,
  Metz2019}.

\begin{acknowledgments}
  The authors acknowledge useful discussions with D.~F.  Agterberg and
  M. Vojta. H.\,M. and P.\,M.\,R.\,B. were supported by the Marsden
  Fund Council from Government funding, managed by Royal Society Te
  Ap\=arangi. H.\,M. acknowledges the hospitality of the Technische
  Universit\"at Dresden. C.\,T.  acknowledges the hospitality of the
  University of Otago and financial support by the Deutsche
  Forschungsgemeinschaft through the Collaborative Research Center
  SFB~1143, Project A04, the Research Training Group GRK~1621, and the
  Cluster of Excellence on Complexity and Topology in Quantum Matter
  ct.qmat (EXC~2147).
\end{acknowledgments}

\appendix

\section{Evaluation of the Ginzburg-Landau free energy}
\label{sec:GL}

The GL free energy in Eq.~\eqref{eq:GL} can be evaluated analytically
using natural approximations.  For a single-band superconductor, it is
usually assumed that the DOS at the Fermi energy is constant such that
the sum over momenta in Eq.~\eqref{eq:GL-F} can be recast into an
integral over energy. We adapt this method to our two-band model by
rewriting the eigenenergies in Eq.~\eqref{eq:EV} in terms of an
``unsplit'' dispersion $\epsilon_0(\bm{k})$ and a cubic form factor
$f(\hat{\bm{k}})$ with unit vector $\hat{\bm{k}}$,
\begin{equation}
  \epsilon_{\bm{k},\pm}
  = \biggl(1 \pm \frac{f(\hat{\bm{k}})}{\alpha + 5\beta/4}\biggr)\,
  \epsilon_0(\bm{k})
  \pm \frac{f(\hat{\bm{k}})}{\alpha + 5\beta/4}\, \mu ,
\end{equation}
where
\begin{align}
  \epsilon_0(\bm{k})
  &= \biggl(\alpha + \frac{5\beta}{4}\biggr)\, \abs{\bm{k}}^2 - \mu, \\
  f(\hat{\bm{k}})
  &= \frac{\beta}{\abs{\bm{k}}^2} \sqrt{\sum_i \biggl[ k_i^4 +
    \biggl( \frac{3\gamma^2}{\beta^2} - 1 \biggr) k_i^2 k_{i+1}^2 \biggr]}.
\end{align}
In the spherical limit $\beta=\gamma$, the form factor reduces to
$f(\theta,\phi) = \beta$, which is angle independent. Then, assuming
constant normal-state DOS, we make the replacement,
\begin{equation}
  \label{eq:GL-approx}
  \sum_{\bm{k}} \to
  \mathcal{N}_0   \int_{\mathbb{S}_2} \frac{\diff\Omega}{4\pi}
    \int_{-\infty}^\infty \diff\epsilon_0 ,
\end{equation}
where $\mathcal{N}_0 = \sqrt{\mu}/2\,(\alpha + 5\beta/4)^{3/2}$ is the
``unsplit'' normal-state DOS at the Fermi energy.

In the case of the fourth-order term, the integral over energy
$\epsilon_0$ and the following summation over $\ii \omega_n$ results
in a sum of polygamma functions which does not yield particular
insight and is not reproduced here.  Nevertheless, below we
demonstrate how to obtain GL coefficients with the outlined approach
for the example of the lowest-order coupling between the
superconducting and the magnetic order parameters.

\section{Third-order term}
\label{sec:F3}

\begin{figure}
  \centering
  \includegraphics{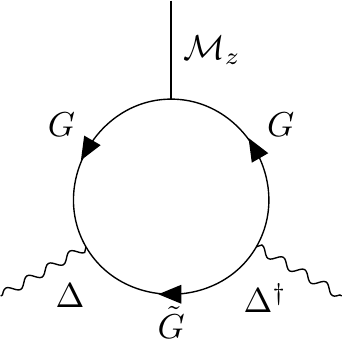}
  \caption{General form of the diagrams that are generated by the
    third-order term of the GL free energy.  Note that $\mathcal{M}_z$
    always connects two lines of the same kind, whereas $\Delta$
    connects to one particle- and one holelike line.}
  \label{fig:F3-diagram}
\end{figure}

In this appendix, we outline the derivation of the leading term in the
GL expansion that couples the superconducting and magnetic order
parameters. In the spherical limit $\beta=\gamma$, the Green's
functions of the particlelike and holelike excitations of the normal
state have the explicit forms
\begin{align}
  G_0(\bm{k}, \ii\omega_n) &= \sum_\pm G_\pm\,
    \frac{1 \pm [(\hat{\bm{k}} \cdot \bm{J})^2 - 5/4]}{2}, \\
  \tilde{G}_0(\bm{k}, \ii\omega_n) &= \sum_\pm \tilde{G}_\pm\,
    \frac{1 \pm [(\hat{\bm{k}} \cdot \bm{J}^T)^2 - 5/4]}{2},
\end{align}
where unit matrices have been suppressed and we have introduced the
single-band Green's functions,
\begin{align}
  G_\pm &\equiv \frac{1}{\ii\omega_n - \epsilon_{\bm{k},\pm}}, \\
  \tilde{G}_\pm &\equiv \frac{1}{\ii\omega_n + \epsilon_{\bm{k},\pm}} .
\end{align}
The magnetic and superconducting order parameters are given by
\begin{align}
  \mathcal{M}_z &= \frac{2}{3}\, m_z (7 J_z - 4 J_z^3) , \\
  \Delta &= \Delta_0 (\eta_{yz} + \ii \eta_{xz}) ,
\end{align}
respectively, and are arranged in the matrix $\Sigma$ as shown in
Eq.~\eqref{eq:Sigma_new}.  With these definitions, the trace in the
third-order coefficient can be expanded in products of $G_\pm$ and
$\tilde{G}_\pm$.  We denote this product without the prefactors as
$F_3$ such that
\begin{equation}
  k_B T \sum_{\bm{k},\omega_n} \frac{1}{3} \tr[(G \Sigma)^3]
    = F_3 m_z\abs{\Delta_0}^2.
\end{equation}
Figure~\ref{fig:F3-diagram} shows the general diagrammatic form of the
generated term for which there are 12 possibilities.  However, four
of these have vanishing coefficients so that only eight terms remain
in two groups of four,
\begin{widetext}
\begin{align}\label{eq:F3greensfunctions}
    F_3 &= k_B T \sum_{\bm{k},\omega_n} \{
    6 \sin^2(2\theta) 
    (
    - G_- \tilde{G}_- G_+ 
    - G_- G_+ \tilde{G}_+ 
    + \tilde{G}_- G_+ \tilde{G}_+ 
    + G_- \tilde{G}_- \tilde{G}_+ 
    ) \notag
    \\
    &\quad {}+
    [5 + 3 \cos(4\theta)] 
    (
    - \tilde{G}_- G_+ G_+ 
    + G_- \tilde{G}_+ \tilde{G}_+ 
    + \tilde{G}_- \tilde{G}_- G_+ 
    - G_- G_- \tilde{G}_+ 
    ) \} ,
\end{align}
where $\theta$ is the polar spherical angle of $\bm{k}$.  There is no
contribution where the Green's functions all have the same band index,
which shows that the coupling to the magnetic order parameter requires
interband pairing.  The combination of Green's functions appearing in
the first line couples the interband component of the magnetic order
parameter to one interband and one intraband component of the
superconducting pairing potential. On the other hand, the combination
of Green's functions in the second line couples the intraband
component of the magnetic order parameter to two interband components
of the superconducting order.  The latter terms correspond to the
coupling of the magnetic order parameter with the pseudomagnetic field
in the low-energy effective model.  Using the approximation from
Eq.~\eqref{eq:GL-approx}, we find that only this term gives a nonzero
contribution,
\begin{align}
    k_B T \sum_{\omega_n} \int_{-\infty}^{\infty} \diff\epsilon_0 \bigl(
    - G_- \tilde{G}_- G_+
    - G_- G_+ \tilde{G}_+
    + \tilde{G}_- G_+ \tilde{G}_+
    + G_- \tilde{G}_- \tilde{G}_+
    \bigr) &= 0, \\
    k_B T \sum_{\omega_n} \int_{-\infty}^{\infty} \diff\epsilon_0 \bigl(
    - \tilde{G}_- G_+ G_+
    + G_- \tilde{G}_+ \tilde{G}_+
    + \tilde{G}_- \tilde{G}_- G_+
    - G_- G_- \tilde{G}_+
    \bigr) &= - \frac{1}{\pi k_B T_c}\, \tilde\beta\, \imag\biggl[
      \psi^{(1)}\biggl(\frac{1}{2}
      + \frac{\ii \tilde\beta \mu}{2 k_B T_c \pi}\biggr)\biggr],
\end{align}
\end{widetext}
where $\psi^{(n)}(z)$ is the polygamma function of order $n$ and
$\tilde\beta = \beta/(\alpha + 5\beta/4)$.  Performing the angular
integration, we obtain
\begin{align}
  F_3 &= -\mathcal{N}_0\, \frac{24}{\pi k_B T_c}\, g_M \abs{\Delta_0}^2
  \tilde\beta \imag\biggl[\psi^{(1)}\biggl(\frac{1}{2}
    + \frac{\ii \tilde\beta
      \mu}{2 \pi k_B T_c}\biggr)\biggr] \notag\\
  &\approx \frac{\mathcal{N}_0}{\mu}\, \frac{48}{5}\, g_M \abs{\Delta_0}^2 ,
\end{align}
where the last approximation is valid when the band splitting
$\tilde{\beta}\mu$ is much larger than $k_BT_c$.

The coefficient $F_3$ is on the order of
$\mathcal{N}_0/\mu \approx \mathcal{N}_0^\prime$, i.e., the derivative
of the DOS at the Fermi energy. This suggests that we should also
include the contributions due to the particle-hole asymmetry of the
normal-state electronic structure, which should also be proportional
to the derivative of the DOS.  To this end, we expand the DOS up to
first order in energy,
$\mathcal{N}(\epsilon_0) \mathrel{\approx} \mathcal{N}_0\, [1 +
\epsilon_0/(2\mu)]$. We have already evaluated the contribution of the
constant term; including the energy-dependent term, however, typically
leads to the divergence of the Matsubara sum. We, therefore, introduce
an energy cutoff such that the sum is restricted to
$\abs{\omega_n} < \Lambda$ where $\Lambda$ is the cutoff energy of the
attractive pairing interaction~\cite{Coleman}. Evaluating the
different sets of Green's functions in
Eq.~\eqref{eq:F3greensfunctions}, we obtain
\begin{widetext}
\begin{align}
& k_B T \sum_{\abs{\omega_n}<\Lambda}
    \int_{-\infty}^\infty \diff\epsilon_0 \: \frac{\epsilon_0}{2\mu}\, \bigl(
    - G_- \tilde{G}_- G_+
    - G_- G_+ \tilde{G}_+
    + \tilde{G}_- G_+ \tilde{G}_+
    + G_- \tilde{G}_- \tilde{G}_+
    \bigr)
  = -\frac{H_{{\frac{\Lambda }{2 k_B T \pi}}}+\ln 4}{\mu (1 - \tilde\beta^2)} , \\
& k_B T \sum_{\abs{\omega_n}<\Lambda}
    \int_{-\infty}^\infty \diff\epsilon_0 \: \frac{\epsilon_0}{2\mu}\, \bigl(
    - \tilde{G}_- G_+ G_+
    + G_- \tilde{G}_+ \tilde{G}_+
    + \tilde{G}_- \tilde{G}_- G_+
    - G_- G_- \tilde{G}_+ \bigr) \nonumber \\
&\quad = \frac{1}{2 \mu } \left(
    2 \real\left[ H_{-\frac{1}{2}+\frac{i \tilde\beta \mu}{2 k_B T \pi }} \right]
    - 2 \real\left[ H_{\frac{i \tilde\beta \mu+\Lambda }{2 k_B T \pi }} \right]
  \right) ,
\end{align}
\end{widetext}
where $H_z$ is the analytic continuation of the harmonic
number. Combining these results with the contribution of the
constant-DOS term, we obtain
\begin{widetext}
  \begin{align}
  F_3 &= \mathcal{N}_0 g_M \abs{\Delta_0}^2\, \frac{24}{5\pi}\, \biggl\{
  - \frac{\tilde\beta}{k_B T_c}
  \imag\biggl[\psi^{(1)}\biggl(\frac{1}{2}
  + \frac{\ii \tilde\beta \mu}{2 \pi k_B T_c}\biggr)\biggr]
  - \frac{2 \pi}{3}\, \frac{H_{\frac{\Lambda }{2 k_B T_c \pi}}+\ln 4}{\mu (1 - \tilde\beta^2)}
  + \frac{\pi}{\mu } \real\left[
    H_{-\frac{1}{2}+\frac{i \tilde\beta \mu}{2 k_B T_c \pi }}
    - H_{\frac{i \tilde\beta \mu+\Lambda }{2 k_B T_c \pi }} \right]
  \biggr\} \notag\\
  &\approx g_M\, \frac{\mathcal{N}_0}{\mu}\, \frac{48}{5} \left[1-
  \frac{\ln\frac{2\Lambda e^\gamma}{\pi
      k_BT_c}}{3(1-\tilde{\beta}^2)}-\frac{1}{4} \ln
      \left(1 + \frac{\Lambda^2}{\tilde{\beta}^2\mu^2}\right)\right] ,
  \end{align}
\end{widetext}
where the second line is valid in the limit $\Lambda$,
$\tilde{\beta}\mu \gg k_BT_c$ and $\gamma$ is the Euler-Mascheroni
constant.


%

\end{document}